\documentclass{article}

\usepackage{PRIMEarxiv}
\usepackage{amsthm, amsmath}
\usepackage[utf8]{inputenc} 
\usepackage[T1]{fontenc}    
\usepackage{hyperref}       
\usepackage{url}            
\usepackage{booktabs}       
\usepackage{amsfonts}       
\usepackage{nicefrac}       
\usepackage{microtype}      
\usepackage{lipsum}
\usepackage{fancyhdr}       
\usepackage{graphicx}       
\graphicspath{{media/}}     
\usepackage{hyperref}
\usepackage{cleveref}
\usepackage{pdflscape}

\title{Experimental implementation of an economic model predictive control for froth flotation\thanks{This work was submitted to and accepted by the 34th European Symposium on Computer Aided Process Engineering \& International Symposium on Process Systems Engineering (ESCAPE34/PSE24) conference, Florence, Italy. July 2024. DOI: \url{https://doi.org/10.1016/B978-0-443-28824-1.50294-5}}}

\author{
  Paulina Quintanilla$^{1}$\thanks{Corresponding author. Current affiliation: Department of Chemical Engineering, Brunel University London, United Kingdom. Email: \texttt{paulina.quintanilla@brunel.ac.uk}}, Daniel Navia$^{2}$, Stephen Neethling$^{1}$, Pablo Brito-Parada$^{1}$ \\
$^1$Department of Earth Science and Engineering, Imperial College London, UK \\
$^2$Departamento de Ingeniería Química y Ambiental, Universidad Técnica Federico Santa María, Chile}

\makeatletter
\input{size11.clo}
\makeatother

\begin{document}
\maketitle

\begin{abstract}
We present the implementation of a novel economic model predictive control (E-MPC) strategy for froth flotation, the largest tonnage mineral separation process. A previously calibrated and validated dynamic model incorporating froth physics was used, which overcomes the limitations of previous simplified models reported in the literature. The E-MPC's optimal control problem was solved using full discretization with orthogonal collocation over finite elements, employing automatic differentiation via CasADi. This approach was applied in a 30-litre laboratory-scale flotation cell, significantly improving mineral recovery from 9\% to 29\% under feed flowrate disturbances while maintaining a minimum concentrate grade of 20\%.

\end{abstract}


\section{Introduction}

The growing demand for minerals and metals in the shift towards cleaner energy sources poses a challenge due to decreasing ore quality. To meet demand effectively, improving the efficiency of current mineral separation techniques while minimizing negative environmental impact is crucial. Froth flotation is the largest tonnage separation process. Given the large-scale nature of this process, even a small improvement in the separation efficiency can result in a substantial increment in mineral recovery \cite{ferreira2000}. 

Effectively controlling the process is difficult due to its complex and dynamic nature \cite{quintanilla2021a} and process disturbances, such as feed flowrate, particle size, and feed grade. The traditional control method used in this process is Proportional-Integral (PI) control, primarily used for regulatory control. However, PI controllers alone are usually ineffective in optimizing key performance indicators, especially under process disturbances, leading to suboptimal outcomes. Advanced control and optimization strategies, particularly Model Predictive Control (MPC), have gained significant attention for improving process performance in froth flotation. MPC uses a dynamic model of the process to predict future behavior and optimize control actions, balancing performance while satisfying constraints. However, despite the potential benefits of MPC strategies in flotation, their full utilization has been hindered by the complexity of modeling process dynamics and instabilities. The kinetic models used in previous studies (e.g., \cite{maldonado2007}; \cite{putz2015}; \cite{riquelme2016}) are insufficient in modeling complex froth phase phenomena, which are critical drivers of the process performance. New advancements in flotation modeling for control that incorporate the froth phase phenomena can be found in \cite{oosthuizen2021}, and \cite{quintanilla2021b, quintanilla2021c}.

Economic model predictive control (E-MPC) is a strategy to optimize control actions based on economic objectives. As such, it is a promising solution for enhancing flotation process efficiencies. E-MPC introduces the economic optimization layer into traditional model predictive control, allowing for direct integration of process economics and feedback control \cite{ellis2014}. This approach considers both technical and economic variables as performance indices and uses nonlinear programming techniques to optimize the set points of control loops.

Our study validates an E-MPC strategy previously developed by \cite{quintanilla2023a}. The strategy was tested in a flotation rig described in \cite{quintanilla2023b}. We used a novel dynamic model incorporating froth physics, which was previously calibrated and experimentally validated. We selected the objective function based on sensitivity analyses, considering air recovery (a measurement of froth stability, directly linked to the flotation performance), separation efficiency (mineral recovery), and product quality (concentrate grade) as a proxy of economic performance.

\section{Materials and methods}

\subsection{Model overview and control variables}

We use a nonlinear, dynamic model developed and experimentally validated by \cite{quintanilla2021a, quintanilla2021b}. This model consists of a system of Differential and Algebraic Equations (DAEs), with a total of $K \times I$ equations and variables, where $K$ is the number of bubble size classes, and $I$ is the number of mineralogical classes. In this study, we assume $K = 10$ and two mineralogical classes ($I = 2$): Chalcopyrite (the valuable mineral) and gangue (waste rock, with physical properties similar to quartz). As shown in \cite{quintanilla2023a}, the model was normalized to enhance solver convergence and robustness. The key control variables are summarized in Table~\ref{tab:variables}.

\begin{table}[ht]
\caption{Variables used in control.}
\centering
\begin{tabular}{c c}
\hline
\textbf{Variable} & \textbf{Definition} \\
\hline
Air recovery ($\alpha$) & Fraction of air fed into the cell that overflows \\
Mass ($M_i$) & Mass of the mineralogical classes $i$ \\
Gas holdup ($\phi_k$) & Volume fraction of gas of each bubble size class $k$ \\
Pulp height ($h_p$) & Pulp level \\
Tails flowrate ($Q_{tails}$) & Flowrate of the tailings stream \\
Grade ($G_{conc}$) & Concentration of valuable minerals in the output stream (concentrate) \\
Recovery ($Rec$) & The proportion of valuable minerals retrieved from the total available \\
\hline
\end{tabular}
\label{tab:variables}
\end{table}

\subsection{E-MPC strategy}

The optimal control problem is formulated as a nonlinear programming (NLP) problem using full discretization with orthogonal collocation over finite elements, implemented using MATLAB R2021B with automatic differentiation via CasADi \cite{andersson2019}. The Interior Point Optimizer (IPOPT) solver was employed to solve the NLP. The sampling time for the optimization was set at 1 second, and each iteration's solution time averaged 0.6 seconds.
\newpage
The general form of the NLP with normalized variables is given by:

\begin{equation}
\begin{aligned}
\min_{\mathbf{\tilde{x}} \in \mathbb{R}^{N_x}, \mathbf{\tilde{z}} \in \mathbb{R}^{N_z}, \mathbf{u_n} \in \mathbb{R}^{N_u} \text{ for } n=0,\ldots,N_p} \ J(\mathbf{\tilde{x}}, \mathbf{y}, \mathbf{\tilde{z}}, \mathbf{u}, \boldsymbol{\beta}, \tilde{t})\\
\text{s.t.} \quad  \mathbf{h}(\mathbf{\tilde{x}}, \mathbf{y}, \mathbf{\tilde{z}}, \mathbf{u}, \boldsymbol{\beta}, \tilde{t}) = 0, \ \tilde{t} \in [t_0, t_{N_p}]\\
    \mathbf{g}(\mathbf{\tilde{x}}, \mathbf{y}, \mathbf{\tilde{z}}, \mathbf{\tilde{u}}, \boldsymbol{\beta}, \tilde{t}) \leq 0, \ \tilde{t} \in [t_0, t_{N_p}]\\
    \mathbf{u}(\tilde{t}) = \mathbf{u_n}, \ \tilde{t} \in [t_n, t_{n+1}] \quad n = 0,\ldots,N_p-1
\end{aligned}
\end{equation}

where the normalized time ($\tilde{t}$) limits are $t_0 = 0$ and $t_{N_p} = 1$, the state vector is $\boldsymbol{\tilde{x}} := \begin{bmatrix} \tilde{M}_i, \tilde{\phi}_k, \tilde{h}_p, \tilde{Q}_{tails} \end{bmatrix}^T$, the process variables vector is $\boldsymbol{\mathbf{y}}:= \begin{bmatrix} \alpha, G_{conc}, Rec \end{bmatrix}^T$, the algebraic variable vector is $\tilde{z}:= \begin{bmatrix} \alpha, \alpha^* \end{bmatrix}^T$, the decision variable vector is $\mathbf{u}_n := \begin{bmatrix} j_g^{SP}, h_p^{SP} \end{bmatrix}^T$ (the superscript $SP$ indicates that the variable is a set point for regulatory controllers), the weight vector is $\boldsymbol{\beta} := \begin{bmatrix} \beta_\alpha, \beta_{G_{conc}}, \beta_{Rec}, \beta_{u_n} \end{bmatrix}^T = \begin{bmatrix} 10^8, 10^6, 10^8, 10^6, 10^2 \end{bmatrix}^T$, $\boldsymbol{h}$ is the dynamic model from \cite{quintanilla2021b, quintanilla2021c}, and $\boldsymbol{g}$ are the process constraints defined in \cite{quintanilla2023a}.

The objective function was selected via sensitivity analyses with respect to the decision variables, as discussed in \cite{quintanilla2023a}, and is defined as:

\begin{equation}
J = \int_{t_0}^{t_{N_p}} \left( \beta_{G_{conc}} G_{conc}(t) - \beta_\alpha \alpha(t) \right) dt - Rec(t_p) + \sum_{n=0}^{N_p-1} (\Delta \mathbf{u}_n^T \boldsymbol{\beta}_u \Delta \mathbf{u}_n),
\label{eq:2}
\end{equation}

where $G_{conc}(t)$ is the dynamic concentrate grade, $\alpha(t)$ is the dynamic air recovery, $Rec(t_{N_p})$ is the mineral recovery at the end of the prediction horizon ($t_{N_p}$), and $\Delta \mathbf{u}_n$ are the decision variable vectors. It is important to note that the metallurgical recovery ($Rec$) is only meaningful at a steady state, which is why it is maximized at the end of the prediction horizon. To balance the trade-off between metallurgical recovery and concentrate grade, we propose to minimize the dynamic concentrate grade in the objective function (\Cref{eq:2}) while imposing a process constraint where $G_{conc}(t) \geq 20\%$ to ensure minimum quality.

\subsection{Experimental setup and implementation}

\begin{figure}[h!]
    \centering
    \includegraphics[width=1\linewidth]{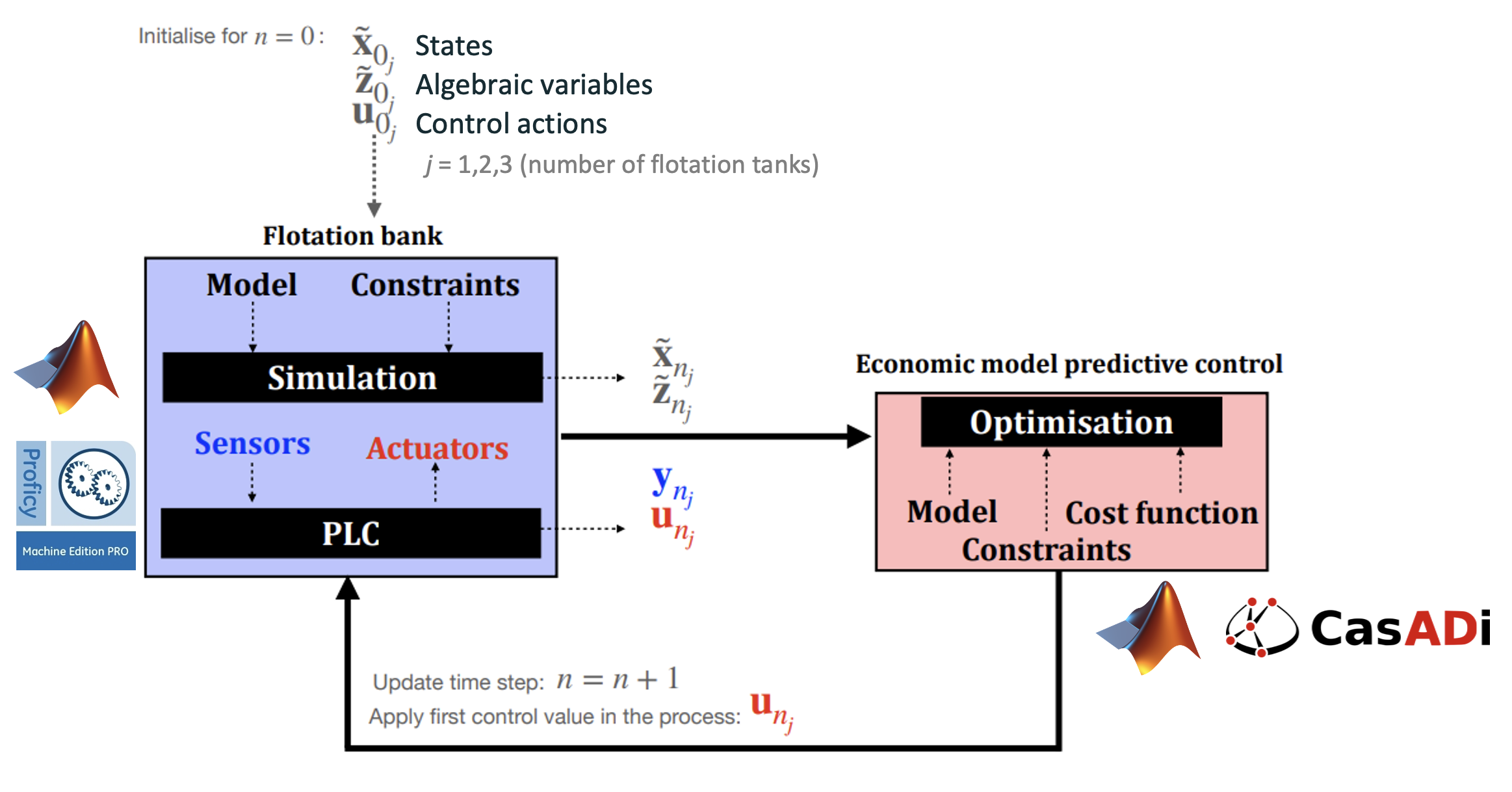}
    \caption{Implementation framework of the E-MPC strategy at the laboratory scale. The E-MPC determines the optimal control actions, which are sent to the Programmable Logic Controller (PLC). The PLC signals the actuators (valves) to reach the set points.}
    \label{fig:empc_framework}
\end{figure}

The experimental setup consists of a 30-litre flotation cell as described in \cite{quintanilla2023b}. The system includes an airflow control system, peristaltic pumps for feed and pulp level regulation, and sensors for real-time measurement of air recovery and pulp level. The instrumentation is linked to the Proficy Machine Edition software, which is connected to MATLAB via an I/O server for real-time data acquisition and control (Figure~\ref{fig:empc_framework}).


The feed flowrate ($Q_{feed}$) is a measurable process disturbance. To test the robustness of the EMPC under disturbances, we performed step changes to the feed flowrate using four different values: 52.5, 56, 63, and 66.5 liters per minute [lpm]. Each feed flowrate was kept constant for 5 minutes, and each iteration was implemented every 1 minute in the laboratory-scale flotation rig.

\newpage
\section{Results and discussions}

\begin{figure}[h!]
    \centering
    \includegraphics[width=0.8\linewidth]{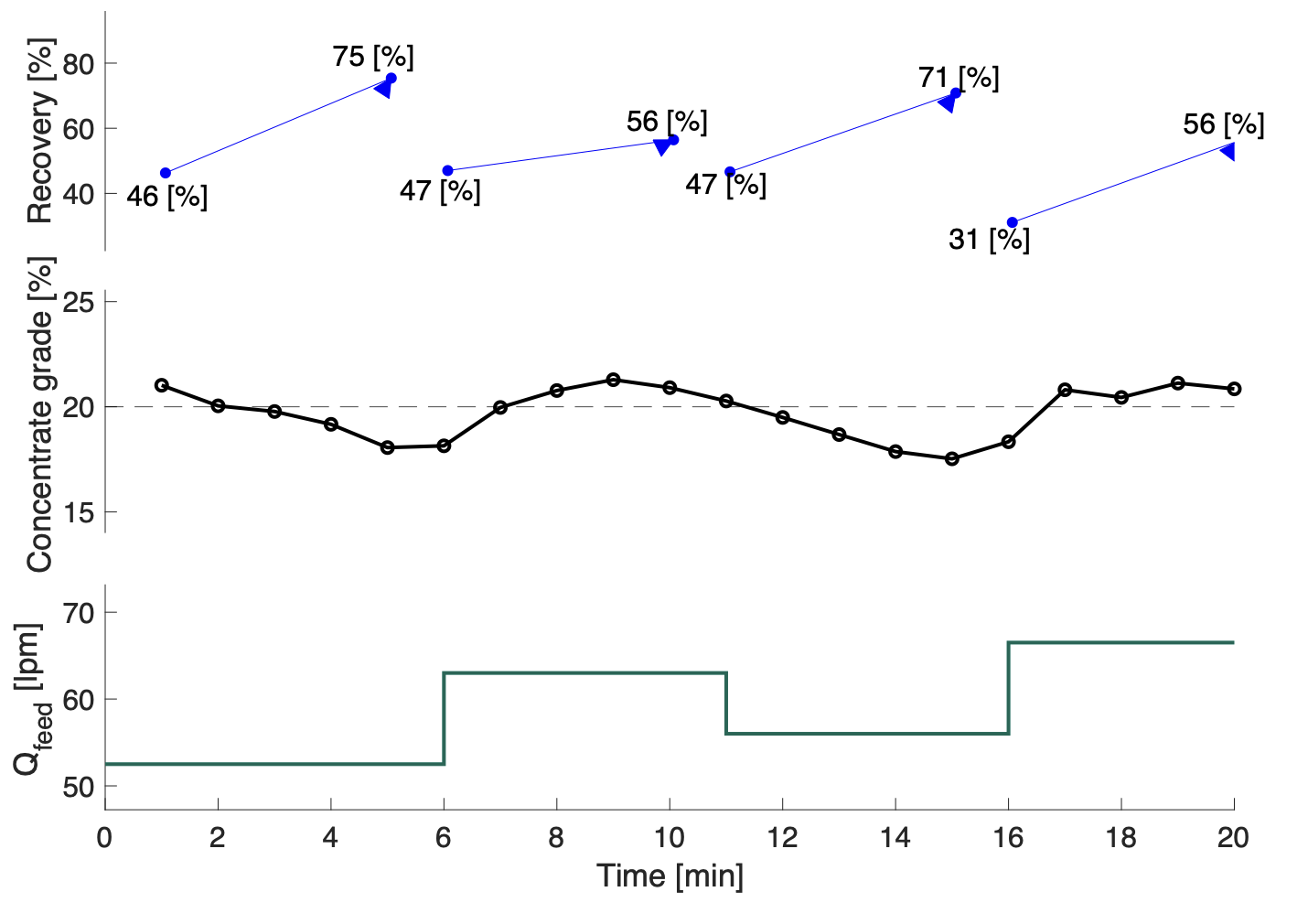}
    \caption{Mineral recovery and concentrate grade for different feed flowrates.}
    \label{fig:results}
\end{figure}

Figure~\ref{fig:results} shows the recovery and concentrate grade as a function of feed flowrate. These experimental results showed that the E-MPC strategy improved mineral recovery between 9\% and 29\%, depending on the feed flowrate conditions. The concentrate grade was maintained near 20\% in most iterations, with some lower grades observed at the lowest feed flowrate, as discussed in \cite{quintanilla2023b}. The discrepancies in control performance were likely due to the differences in the Proportional-Integral (PI) controller's sampling time and the model-based E-MPC system.


\begin{figure}[h!]
    \centering
    \includegraphics[width=0.8\linewidth]{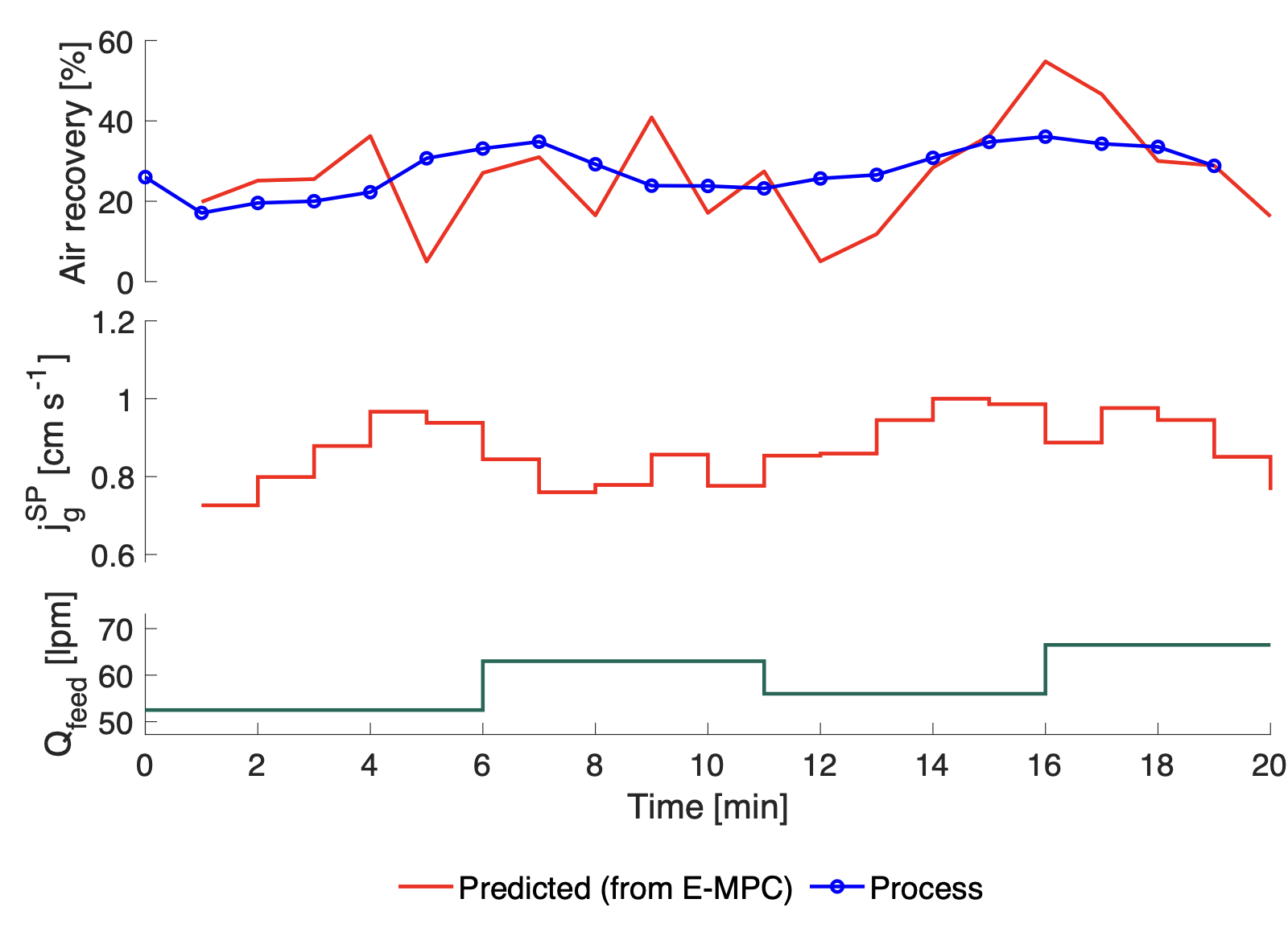}
    \caption{Air recovery and superficial air velocity for changes in feed flowrate.}
    \label{fig:air_recovery}
\end{figure}

\begin{figure}[h!]
    \centering
    \includegraphics[width=0.8\linewidth]{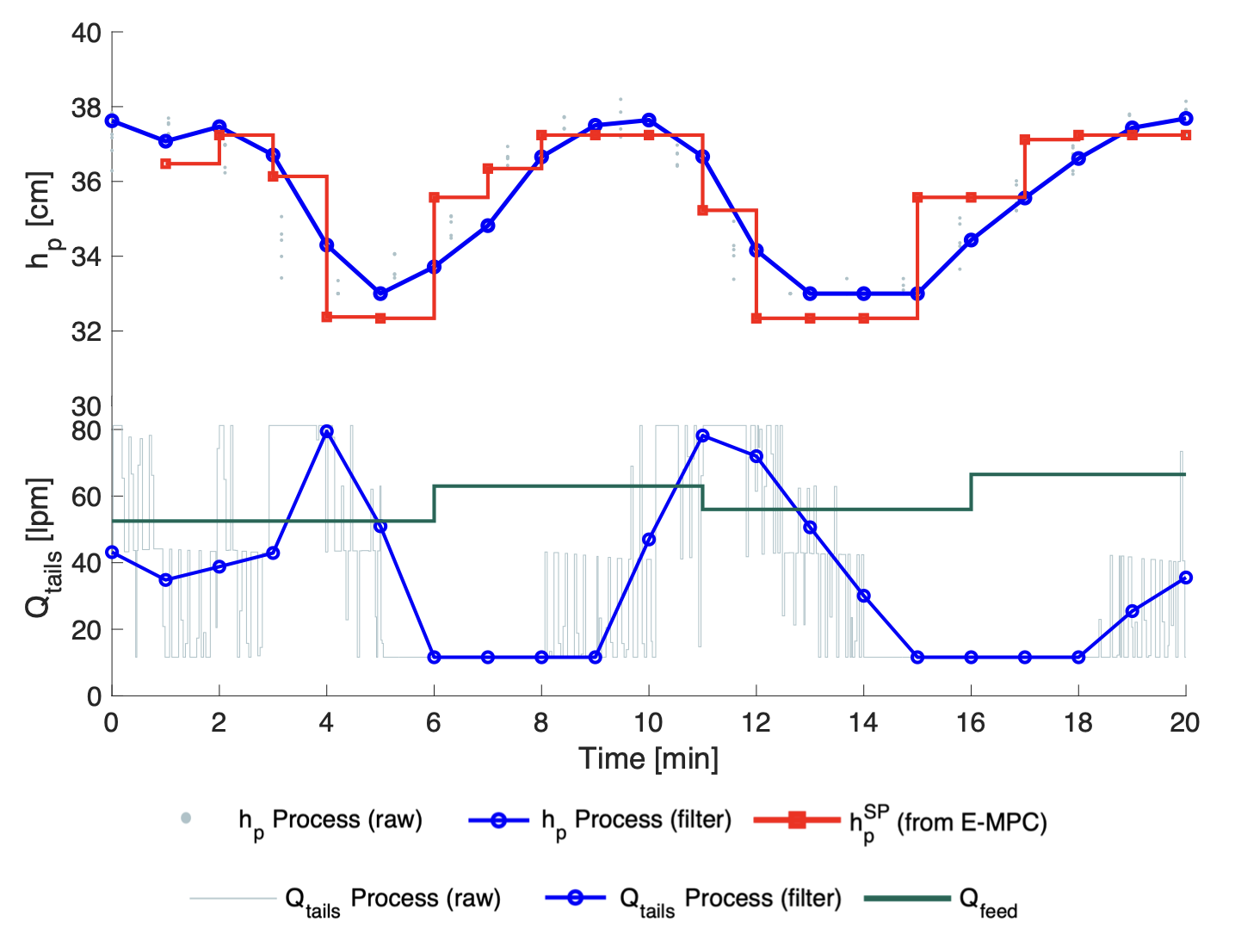}
    \caption{Level control using tail flowrates($Q_{tails}$). Red lines are set points from E-MPC optimization, and blue lines are filtered pulp height ($h_p$) in process.}
    \label{fig:level_control}
\end{figure}

As shown in Figure~\ref{fig:air_recovery}, air recovery generally followed the trends predicted by the E-MPC strategy, given the different values of superficial air velocity. Moreover, it can be observed that the highest air recoveries were obtained for the highest values of feed flowrates (63 and 66.5 [lpm]), which coincides with the highest pulp heights (see Figure~\ref{fig:level_control}). According to what was observed during the experiments, the increase in air recovery may be related to increments in overflowing froth velocity due to shallower froth depths.


Figure~\ref{fig:level_control} shows the level control for the different pulp height setpoints sent from the E-MPC. While the trends of the pulp heights in the systems are the same as the setpoints, the values are usually different, which may be related to the differences between the parameters of the Proportional-Integral (PI) in the laboratory-scale system and the model used for the E-MPC. Those parameters are different because the sampling times are not the same in both cases, i.e. the PI controller had a sampling time of 1 second, while the model used in the E-MPC strategy corresponded to 10 seconds


\section{Conclusions}
This study validates an E-MPC strategy in a laboratory-scale flotation cell using a novel dynamic model that considers froth physics. The results of the experiments reveal that the E-MPC approach has led to higher metallurgical recoveries and has also shown its potential to handle feed flowrate disturbances efficiently. Further research will validate the E-MPC strategy in a laboratory-scale flotation bank (i.e. several tanks interconnected in series) to mimic the systems found in industrial-scale operations.

\bibliographystyle{unsrt}  
\bibliography{references}

\begin{thebibliography}{10}

\bibitem{ferreira2000}
J.~P. Ferreira and B.~K. Loveday.
\newblock An improved model for simulation of flotation circuits.
\newblock {\em Minerals Engineering}, 13(14-15):1441--1453, 2000.

\bibitem{quintanilla2021a}
P.~Quintanilla, S.~J. Neethling, and P.~R. Brito-Parada.
\newblock Modelling for froth flotation control: A review.
\newblock {\em Minerals Engineering}, 162:106718, 2021.

\bibitem{maldonado2007}
M.~Maldonado, D.~Sbarbaro, and E.~Lizama.
\newblock Optimal control of a rougher flotation process based on dynamic programming.
\newblock {\em Minerals Engineering}, 20(3):221--232, 2007.

\bibitem{putz2015}
E.~Putz and A.~Cipriano.
\newblock Hybrid model predictive control for flotation plants.
\newblock {\em Minerals Engineering}, 70:26--35, 2015.

\bibitem{riquelme2016}
A.~Riquelme, A.~Desbiens, R.~Del Villar, and M.~Maldonado.
\newblock Predictive control of the bubble size distribution in a two-phase pilot flotation column.
\newblock {\em Minerals Engineering}, 89:71--76, 2016.

\bibitem{oosthuizen2021}
D.~J. Oosthuizen, J.~D. le~Roux, and I.~K. Craig.
\newblock A dynamic flotation model to infer process characteristics from online measurements.
\newblock {\em Minerals Engineering}, 167:106878, 2021.

\bibitem{quintanilla2021b}
P.~Quintanilla, S.~J. Neethling, D.~Navia, and P.~R. Brito-Parada.
\newblock A dynamic flotation model for predictive control incorporating froth physics. part i: Model development.
\newblock {\em Minerals Engineering}, 173:107192, 2021.

\bibitem{quintanilla2021c}
P.~Quintanilla, S.~J. Neethling, D.~Mesa, D.~Navia, and P.~R. Brito-Parada.
\newblock A dynamic flotation model for predictive control incorporating froth physics. part ii: Model calibration and validation.
\newblock {\em Minerals Engineering}, 173:107190, 2021.

\bibitem{ellis2014}
M.~Ellis, H.~Durand, and P.~D. Christofides.
\newblock A tutorial review of economic model predictive control methods.
\newblock {\em Journal of Process Control}, 24(8):1156--1178, 2014.

\bibitem{quintanilla2023a}
P.~Quintanilla, D.~Navia, S.~J. Neethling, and P.~R. Brito-Parada.
\newblock Economic model predictive control for a rougher froth flotation cell using physics-based models.
\newblock {\em Minerals Engineering}, 196:108050, 2023.

\bibitem{quintanilla2023b}
P.~Quintanilla, D.~Navia, F.~Moreno, S.~J. Neethling, and P.~R. Brito-Parada.
\newblock A methodology to implement a closed-loop feedback-feedforward level control in a laboratory-scale flotation bank using peristaltic pumps.
\newblock {\em MethodsX}, 10:102081, 2023.

\bibitem{andersson2019}
J.~A.~E. Andersson, J.~Gillis, G.~Horn, J.~B. Rawlings, and M.~Diehl.
\newblock Casadi: a software framework for nonlinear optimization and optimal control.
\newblock {\em Mathematical Programming Computation}, 11:1--36, 2019.

\end{thebibliography}

\end{document}